\documentclass[
    ,final            
    ,numberedheadings 
  ]
  {aipproc}

\usepackage{bm}
\usepackage{epsfig}
\layoutstyle{6x9}

\def\HeII{He\,{\footnotesize II}}

\begin{document}

\title{On the seismic age of the Sun}



\author{G. Houdek}{
address={Institute of Astronomy, Madingley Road, Cambridge CB3\,0HA, UK}
}
\author{D. O. Gough$^{*}$}{
address={Department of Applied Mathematics and Theoretical Physics, 
                               Wilberforce Road, Cambridge CB3\,0WA, UK} 
}

\begin{abstract}
We use low-degree acoustic modes obtained by the BiSON  to estimate 
the main-sequence age $t_\odot$ of the Sun. The calibration is accomplished 
by linearizing the deviations from a standard solar model the seismic 
frequencies of which are close to those of the Sun.  Formally, we obtain the 
preliminary value $t_\odot=4.68\pm0.02\,$Gy, coupled with an initial 
heavy-element abundance $Z=0.0169\pm0.0005$.  The quoted standard errors, 
which are not independent, are upper bounds implied under the assumption that
the standard errors in the observed frequencies are independent. 
\end{abstract}

\maketitle


\section{Introduction}
Seismological calibration of stellar models against observed 
frequencies of low-degree modes was first discussed more than
two decades ago \citep{jcd84, jcd88, ulrich86, dog87}, and can be 
regarded as a means of determining the main-sequence age of the Sun 
\citep{guenther89, dog_nov90, 
guenther_demarque97, 
weiss_schlattl98, wd99, dog01, bon_schlat_pat02}. 
The procedure is to match certain appropriate seismic signatures 
of theoretical frequencies determined on a grid of stellar 
models with corresponding signatures obtained from the 
observations.  The signatures are chosen to reflect principally the 
properties of the energy-generating core, where nuclear 
transmutation leaves behind an augmenting concentration of
helium, lowering the sound speed relative to the environs and 
thereby providing a diagnostic of age.  But the signatures are also 
susceptible to other properties of the stellar interior, which 
must be eliminated before a robust outcome can be achieved.  For 
example, although the so-called small frequency separation is 
sensitive predominantly to the evolving stratification of the core, 
its dependence on the zero-age chemical abundances plays a 
significant contaminating role.  Therefore it behoves us to seek 
an additional diagnostic to attempt to measure abundance separately. 
For given relative 
abundances of the heavy elements, the total absolute heavy-element 
abundance $Z$ and the $^4\rm{He}$ abundance $Y$ are related by 
the requirement that the model has the observed luminosity and radius, 
principally the former.  Therefore we need 
to   
aim at detecting only one of them. 
Here we use a signature indicative of the abrupt variation of the first
adiabatic exponent $\gamma_1$ induced by the ionization of helium. 

\begin{figure}
\includegraphics[height=.42\textheight]{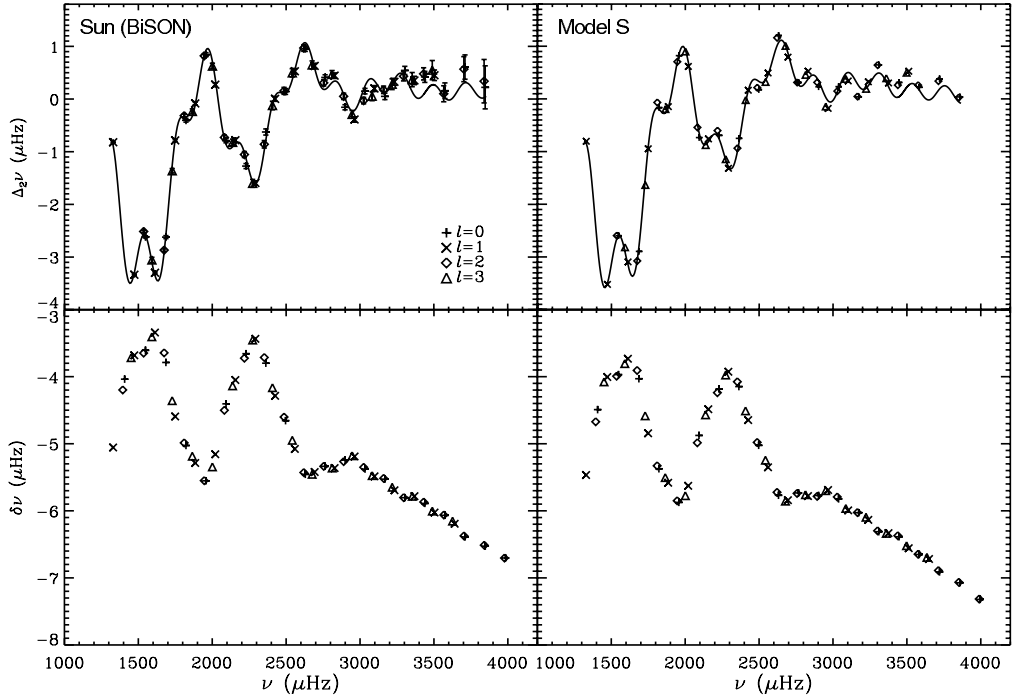}
\caption{Top left: The symbols (with error bars obtained 
under the assumption that the raw frequency errors are independent) 
represent second differences, $\Delta_2\nu$, of low-degree solar
frequencies from BiSON.
Top right: The symbols are second differences $\Delta_2\nu$
of adiabatic pulsation eigenfrequencies of solar 
Model S.
The solid curve in both panels is the diagnostic\,
(\ref{eq:delnu}) -- (\ref{eq:secdiff}), 
whose eleven parameters have been adjusted to fit the data optimally.
Bottom: The symbols denote contributions $\delta\nu$ to the frequencies  
produced by the acoustic glitches 
of the Sun (left panel) and Model S (right panel).}
\end{figure}

\section{The seismic diagnostic and calibration method}
\vspace{-2mm}
Any abrupt variation in the stratification of a star (relative to the scale of 
the inverse radial wavenumber of a seismic mode of oscillation), which here
we call an acoustic glitch, induces an oscillatory component in the spacing of 
the cyclic eigenfrequencies $\nu_{n,l}$ of seismic modes, of order $n$ and 
degree $l$.  Our interest is principally in the glitch caused by the 
depression in the first adiabatic exponent 
$\gamma_1=(\partial {\ln p}/\partial{\ln\rho})_s$ (where $p$, $\rho$ and $s$ 
are pressure, density and specific entropy) caused by helium ionization.
The deviation 
\vspace{-2mm}
\begin{equation}
\delta\nu:=\nu - \nu_{\rm s}
\end{equation}
(from now on we omit the subscripts $n, l$) 
of the eigenfrequency from the corresponding frequency $\nu_{\rm s}$ 
of a similar smoothly stratified star is the indicator of $Y$ that 
we use in conjunction with the indicators of core structure to determine the 
main-sequence age. 

Approximate expressions for the frequency contributions $\delta\nu$ arising 
from acoustic glitches in solar-type stars were recently 
presented by \citet{hgdog07}.  Here we improve them by adopting the 
appropriate Airy functions Ai$(-x)$ that are used as comparison 
functions in the JWKB approximations to the oscillation eigenfunctions.  
The complete expression for $\delta\nu$ is then given by 
\begin{equation}
\delta\nu=\delta_\gamma\nu+\delta_{\rm c}\nu\, ,
\label{eq:delnu}
\end{equation}
\vspace{-2mm}
where
\vspace{-2mm}
\begin{eqnarray}
\delta_\gamma\nu&=&-\sqrt{2\pi}A_{\rm II}\Delta^{-1}_{\rm II}
\left[\nu+\textstyle\frac{1}{2}(m+1)\nu_0\right]\cr
&&\hspace{-8pt}
\times\Bigl[\mu\beta\int_0^T\kappa^{-1}_{\rm I}
{\rm e}^{-(\tau-\eta\tau_{\rm II})^2/2\mu^2\Delta^2_{\rm II}}|x|^{1/2}
|{\rm Ai}(-x)|^2\,{\rm d}\tau\cr
&&\;\;+\;\int_0^T\kappa^{-1}_{\rm II}
{\rm e}^{-(\tau-\tau_{\rm II})^2/2\Delta^2_{\rm II}}|x|^{1/2}
|{\rm Ai}(-x)|^2\,{\rm d}\tau\Bigr]
\label{eq:delgamnu}
\end{eqnarray}
\vspace{-1mm}
arises from the variation in $\gamma_1$ induced by helium ionization, and
\vspace{-1mm}
\begin{eqnarray}
\delta_{\rm c}\nu\!\!&\simeq&\!\!A_{\rm c}\nu_0^3\nu^{-2}
   \left(1+1/16\pi^2\tau_0^2\nu^2\right)^{-1/2}\cr
&\times&\hspace{-8pt}\left\{\cos[2\psi_{\rm c}+\tan^{-1}(4\pi\tau_0\nu)]
      \!-\!(16\pi^2\tilde{\tau}_{\rm c}^2\nu^2\!+\!1)^{1/2}
\right\}\, 
\label{eq:delcnu}
\end{eqnarray}
results from the acoustic glitch at the base of the convection zone. Here, 
$m=3.5$ is a constant, being a representative polytropic index in
the expression for the approximate effective phase $\psi$ appearing in the 
argument of the Airy function, and $\beta$, $\eta$ and $\mu$ are
constants of order unity which account for the relation between the acoustic 
glitches caused by the first and second stages of ionization of helium 
\citep{hgdog07}; $\tau$ is acoustic depth beneath the seismic surface of 
the star, and 
$T=1/2\nu_0$ is the total acoustic radius of the star; $\tau_{\rm II}$ and 
$\Delta_{\rm II}$ are respectively the centre and the width of the 
\HeII\ acoustic glitch.  The 
argument of the Airy function is $x={\rm sun}(\psi)|3\psi/2|^{2/3}$, where 
$\psi(\tau)=\kappa\omega\tilde\tau-(m+1)\cos^{-1}[(m+1)/\omega\tilde\tau]$ 
if $\tilde\tau>\tau_{\rm t}$, and 
$\psi(\tau)=|\kappa|\omega\tilde\tau-(m+1)\ln[(m+1)/\omega\tilde\tau+|\kappa|]$
if $\tilde\tau\le\tau_{\rm t}$, in which 
$\tilde\tau=\tau+\omega^{-1}\epsilon_{\rm II}$, with $\omega=2\pi\nu$, and 
$\tau_{\rm t}$ is the location of the upper turning point;
$\kappa(\tau)=[1-(m+1)^2/\omega^2\tilde\tau^2]^{1/2}$, and
$\kappa_{\rm I}=\kappa(\eta\tau_{\rm II})$ and 
$\kappa_{\rm II}=\kappa(\tau_{\rm II})$.  Also 
$\psi_{\rm c}=\kappa_{\rm c}\omega\tilde\tau_{\rm c}
-(m+1)\cos^{-1}\left[(m+1)/\tilde\tau_{\rm c}\omega\right]+\pi/4$, where 
$\kappa_{\rm c}=\kappa(\tau_{\rm c})$ and 
$\tilde\tau_{\rm c}=\tau_{\rm c}+\omega^{-1}\epsilon_{\rm c}$.
The seven coefficients 
$A_{\rm II}$, $\Delta_{\rm II}$, $\tau_{\rm II}$, $\epsilon_{\rm II}$, 
$A_{\rm c}$, $\tau_{\rm c}$, $\epsilon_{\rm c}$ are found by fitting
the second difference 
\vspace{-3mm}
\begin{equation}
\Delta_2\nu\equiv\nu_{n-1,l}-2\nu_{n,l}+\nu_{n+1,l}
\simeq\Delta_2(\delta_\gamma\nu+\delta_{\rm c}\nu)+\sum_{k=0}^3a_k\nu^{-k}
\label{eq:secdiff}
\vspace{-2mm}
\end{equation}
to the corresponding observations under the assumption that the errors in 
the frequency data are independent (see top panels of Figure\,1). The
last term in equation\,(\ref{eq:secdiff}) approximates smooth contributions
arising, in part, from wave 
refraction in the stellar core, from hydrogen ionization and from the 
superadiabaticity of the upper boundary layer of the convection zone, 
introducing four more fitting coefficients $a_k$ ($k=$0,...,3).

\begin{table}
\centering
\mbox{
\begin{minipage}[h]{70mm}
\begin{tabular}{llll}
\hline
&\hfil$A$\hfil&\hfil$C$\hfil&\hfil$-\delta\gamma_1/\gamma_1$\hfil\\
\hline
\,Sun (BiSON)\,&\,0.2764\,&\,1.785\,&\,0.04325\,\\
\,Model S\,    &\,0.2780\,&\,1.818\,&\,0.04511\,\\
\noalign{\smallskip}
\hline
\end{tabular}
\end{minipage}
\hspace{5mm}
\begin{minipage}[h]{68mm}
  \vspace{-10mm}
  \baselineskip=0.85\normalbaselineskip
  {\AIPtablecaptiontextfont{\bf TABLE 1.\,\ }
   Asymptotic fitting coefficients $A$, $C$ (see equation\,\ref{eq:asymp})
   and $-\delta\gamma_1/\gamma_1=A_{\rm II}/\sqrt{2\pi}\nu_0\Delta_{\rm II}$.
  }
\end{minipage}
}
\stepcounter{table}
\end{table}

The outcome of the fitting to the BiSON data \citep{basu07} and to the 
adiabatically computed eigenfrequencies of solar  Model S \citep{jcd96} is
displayed in Figure\,1: the upper panels display the second differences, 
together with the fitted formula (\ref{eq:secdiff}), the lower panels 
display the corresponding contributions $\delta\nu$ to the frequencies of 
oscillation from the acoustic glitches.
\begin{figure}
\centering
\mbox{
\begin{minipage}[h]{63mm}
 \epsfig{file=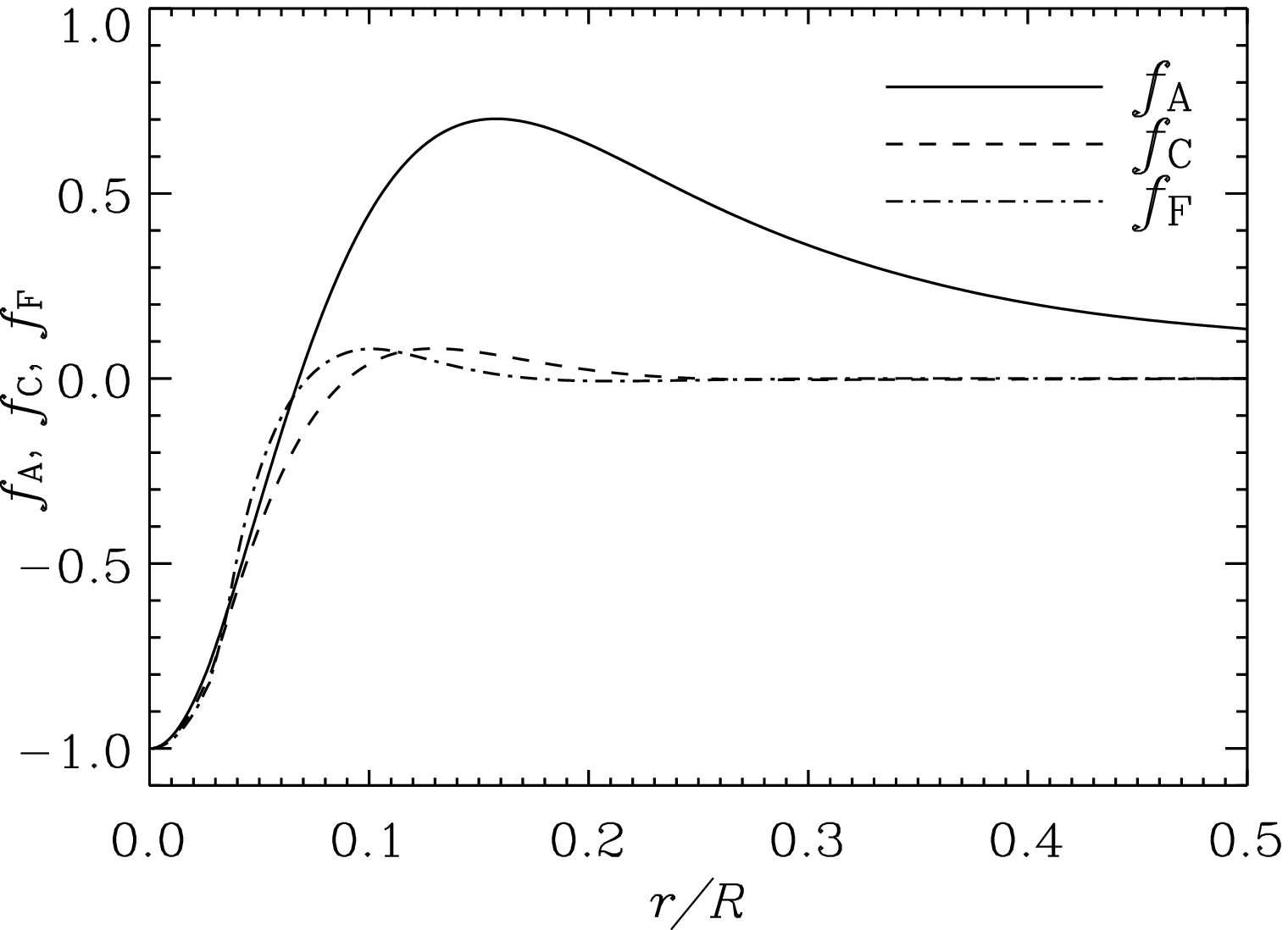,height=.20\textheight}
\end{minipage}
\hspace{2mm}
\begin{minipage}[h]{80mm}
  \vspace{0mm} 
  \baselineskip=0.85\normalbaselineskip
  {\AIPtablecaptiontextfont{\bf FIGURE 2.\,\ }
   Functional forms $f_X$ of the integrands $\phi_X$ in 
   $X=\int_0^R \phi_X {\rm d}r$, where $f_X(r) = \phi_X(r)/|\phi_X(0)|$ 
   and where $X =$\, $A$, $C$, or $F$, plotted over the inner half of 
   the interval $(0,R)$ of $r$. The parameters $A$, $C$ and $F$ are 
   sensitive particularly to the structure of the core, being progressively 
   more centrally concentrated.   For the calibrations here we use only $A$ 
   and $C$ (for $F$ is more difficult to determine with confidence), whose  
   values determined from the fit are listed in Table\,1, together with
   the implied maximum depression 
   $-\delta\gamma_1/\gamma_1=A_{\rm II}/\sqrt{2\pi}\nu_0\Delta_{\rm II}$ in  
   $\gamma_1$ caused by \HeII\ ionization.
  }
\end{minipage}
}
\stepcounter{figure}
\end{figure}
To the resulting smooth (glitch free) frequencies $\nu_{\rm s}$, derived from 
equation (1), of both the solar observations \citep{basu07} and the 
eigenfrequencies of the reference solar model (Model S) was fitted the 
asymptotic expression
\vspace{-2mm}
\begin{equation}
\nu_{n,l}\!\sim\!(n+{\textstyle\frac{1}{2}}\,l+\hat\epsilon)\nu_0
-\frac{A\,L^2\!-\!B}{\nu_{n,l}}\,\nu^2_0
-\frac{C\,L^4\!-\!D\,L^2\!+\!E}{\nu_{n,l}^3}\,\nu^4_0
-\frac{FL^6\!-\!GL^4\!+\!HL^2\!-\!I}{\nu_{n,l}^5}\,\nu^6_0\,,
\label{eq:asymp}
\vspace{-2mm}
\end{equation}
where $L^2=l(l+1)$, from which we obtain the 
coefficients $\nu_0, ~\hat\epsilon,~A, B, C, D, E, F, G, H$ and $I$,  
each of which is an integral of a function of the equilibrium stratification,
some of which are displayed in Figure\,2.    The differences between the 
actual smoothed frequencies $\nu_{\rm s}$ and the asymptotic expression 
(\ref{eq:asymp}) are plotted in Figure\,3. 

\noindent
We have carried out age calibrations using combinations of the parameters 
\begin{equation}
\xi_\alpha=(A,C,-\delta\gamma_1/\gamma_1),\qquad \alpha=1,2,3\,.  
\label{eq:xi}
\end{equation}
Presuming, as is normal, that Model S is parametrically close to the Sun, we 
consider the solar value $\xi^\odot_\alpha$ to be approximated by a two-term 
Taylor expansion of $\xi_\alpha$ about the value $\xi^{\rm s}_\alpha$ for 
Models S:
\vspace{-3mm}
\begin{equation}
\xi^\odot_\alpha=\xi^{\rm s}_\alpha
   +\left(\frac{\partial\xi_\alpha}{\partial t_\odot}\right)_{\!\!Z}\Delta\,t_\odot
   +\left(\frac{\partial\xi_\alpha}{\partial Z}\right)_{\!\!t_\odot}\Delta Z
   -\epsilon_{\xi\alpha}\, ,
\vspace{-1mm}
\end{equation}
where $\Delta\,t_\odot$ and $\Delta Z$ are the deviations of age $t_\odot$ and
initial heavy-element abundance $Z$ from Model\,S, and $\epsilon_{\xi\alpha}$
are the formal errors in the calibration parameters. A (parametrically local) 
maximum-likelihood fit (again, assuming that the errors in the observed 
frequencies are independent) then leads to the following set of linear 
equations: 
\begin{equation}
H_{\alpha j}C^{-1}_{\alpha\beta}H_{\beta k}\Theta_{0k}=
H_{\alpha j}C^{-1}_{\alpha\beta}\Delta_{0\beta}\,,
\label{eq:calib1}
\end{equation}
in which $\Theta_k=(\Delta t_\odot, \Delta Z)+\epsilon_{\Theta k}=
\Theta_{0k}+\epsilon_{\Theta k}$, $k=1,2$ ~is the solution vector subject 
to (correlated) errors $\epsilon_{\Theta k}$, 
$\Delta_\beta=\xi^\odot_\beta-\xi^{\rm s}_\beta+\epsilon_{\xi\beta}
=\Delta_{0\beta}+\epsilon_{\xi\beta}$, $C_{\alpha\beta}$ is the 
covariance matrix of the errors $\epsilon_{\xi\alpha}$, and 
$H_{\alpha j}=
[(\partial\xi_\alpha/\partial t)_Z,(\partial\xi_\alpha/\partial Z)_t]$,
$j=1,2$.
\begin{figure}
  \includegraphics[height=.25\textheight]{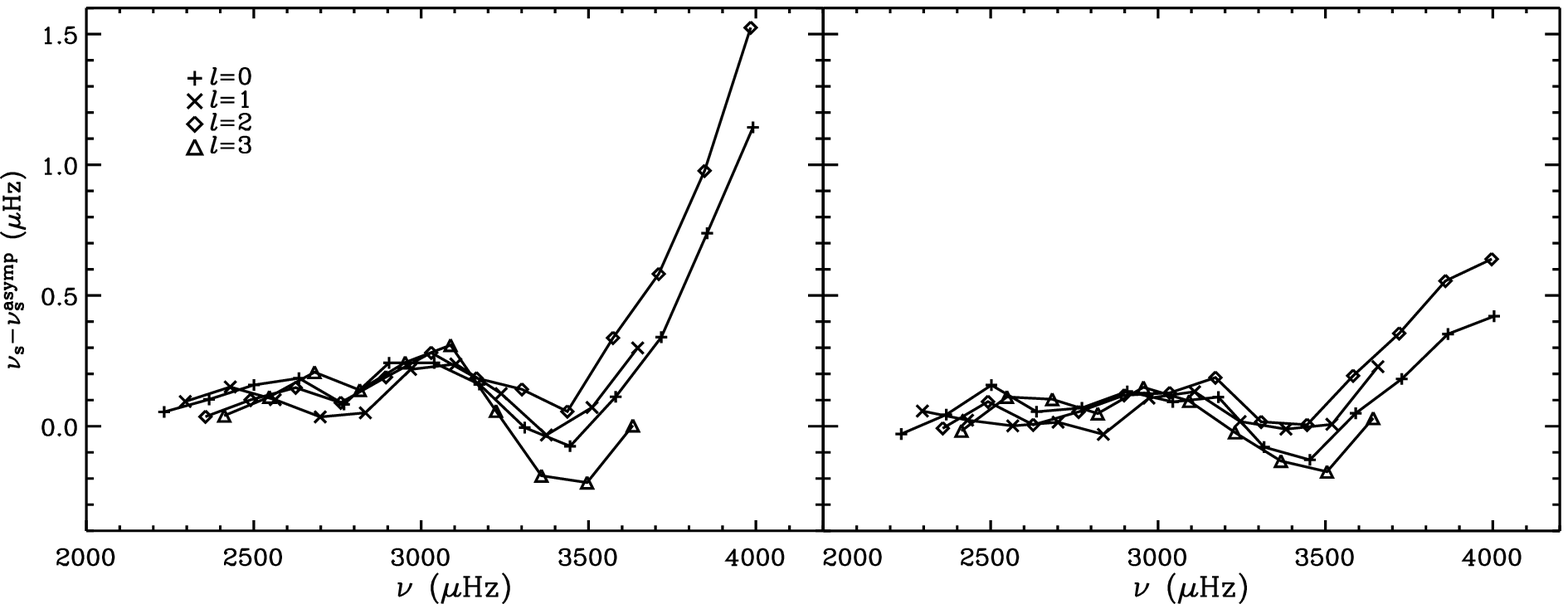}
  \caption{Differences between the smoothed frequencies $\nu_{\rm s}$ 
           of the Sun (left) and Model S (right), and the fitted 
           asymptotic expression (\ref{eq:asymp}). Modes of like 
           degree $l$ are connected by solid lines.}
\end{figure}

\noindent
A similar set of equations is obtained for the formal errors 
$\epsilon_{\Theta k}$: 
\begin{equation}
H_{\alpha j}C^{-1}_{\alpha\beta}H_{\beta k}\epsilon_{\Theta k}=
H_{\alpha j}C^{-1}_{\alpha\beta}\epsilon_{\xi\beta}\,,
\label{eq:calib2}
\end{equation}
from which the error covariance matrix 
$C_{\Theta kq}=\overline{\epsilon_{\Theta k}\epsilon_{\Theta q}}$ can be 
computed with a Monte Carlo simulation. 

\noindent
The partial derivatives $H_{\alpha j}$ are obtained from 
two sets of five calibrated evolutionary models for the Sun,
computed with the evolutionary programme by \citet{jcd82}, and  
adopting the Livermoore equation of state and the OPAL92 opacities. 
One set of models has a constant value for the heavy-element abundance 
$Z=0.02$ but varying age; the other has constant age but varying $Z$. 
The values of the partial derivatives $H_{\alpha j}$ are listed in Table\,2.
\renewcommand\AIPtablesourceheadtext  {}
\begin{table}
\caption{Partial derivatives $H_{\alpha j}$ obtained from two sets of
calibrated evolutionary models for the Sun. Values with respect
to age $t_\odot$ are in units of Gy$^{-1}$.}
\begin{tabular}{cccccc}
\noalign{\smallskip}
\noalign{\smallskip}
\hline
 \hfil$(\partial A/\partial t_\odot)_Z$\hfil&\hfil$(\partial A/\partial Z)_{t_\odot}$\hfil&
 \hfil$(\partial C/\partial t_\odot)_Z$\hfil&\hfil$(\partial C/\partial Z)_{t_\odot}$\hfil&
 \hfil$[\partial(-\delta\gamma_1/\gamma_1)/\partial t_\odot]_Z$\hfil&
 \hfil$[\partial(-\delta\gamma_1/\gamma_1)/\partial Z]_{t_\odot}$\hfil\\
\hline
\,-0.0469\,&\,-0.584\,&\,0.677\,&\,36.8\,&\,-0.00656\,&\,0.442\,\\
\noalign{\smallskip}
\hline
\end{tabular}
\source{\vspace{-5mm}}
\end{table}
\renewcommand\AIPtablesourceheadtext  {Source:~ } 
\vspace{-5mm}
\section{Results}
\vspace{-2mm}
Age calibrations using the different combinations of the parameters 
$\xi_\alpha$ are summarized in Table\,3;  error contours associated with 
the first entry are plotted in Figure\,4.  In all cases the age found is 
greater than currently accepted values.  The values of $Z$ should not be 
regarded strictly as statements about the initial heavy-element abundance, 
but rather as measures of the opacity in the radiative interior. 
\citet{asplund04} have argued that the photospheric abundances of C, N 
and O had previously been overestimated, suggesting that the actual total 
heavy-element abundance is rather lower than previously believed.  However, 
that cannot imply that the opacity in the solar interior is necessarily 
comparably lower because it has been implicitly calibrated here (by accepting 
the tenets of solar-evolution theory, and the OPAL opacity calculations upon 
which the models are based), and indeed the opacity has already been 
determined seismologically from a broader spectrum of modes than has been 
adopted here \citep{dog04}.  The matter raised by Asplund et\,al. therefore 
challenges either the opacity calculations, 
the nuclear reaction rates, or the basic physics of stellar evolution, not 
helioseismology, as some spectators have surmised.   As we 
know already from seismological structure inversions, the solar models are not 
accurate by helioseismological standards.  Therefore the properties inferred 
from these calibrations could be more contaminated by systematic error than by
errors in the observed frequencies.

\begin{table}
\caption{Age calibrations with different combinations of $\xi_\alpha$. 
}
\begin{tabular}{lccccc}
\noalign{\smallskip}
\noalign{\smallskip}
\hline
 \hfil$\xi_\alpha$\hfil&
 \hfil $t_\odot$ (Gy)\hfil&
 \hfil$C^{1/2}_{\Theta 11}$\hfil&
 \hfil$Z$\hfil&
 \hfil$C^{1/2}_{\Theta 22}$\hfil&
 \hfil$-(-C_{\Theta 12})^{1/2}$\hfil\\
\hline
$A,C,-\delta\gamma_1/\gamma_1$&4.679&0.017&0.0169&0.0005&-0.0023\\
$A,C                         $&4.658&0.023&0.0177&0.0007&-0.0037\\
$A,-\delta\gamma_1/\gamma_1  $&4.673&0.017&0.0165&0.0007&-0.0019\\
$C,-\delta\gamma_1/\gamma_1  $&4.700&0.028&0.0169&0.0005&-0.0029\\
\noalign{\smallskip}
\hline
\end{tabular}
\end{table}

\renewcommand\AIPfigurecaptionheadformat[1] {}
\begin{figure}
\centering
\mbox{
\begin{minipage}[h]{63mm}
 \epsfig{file=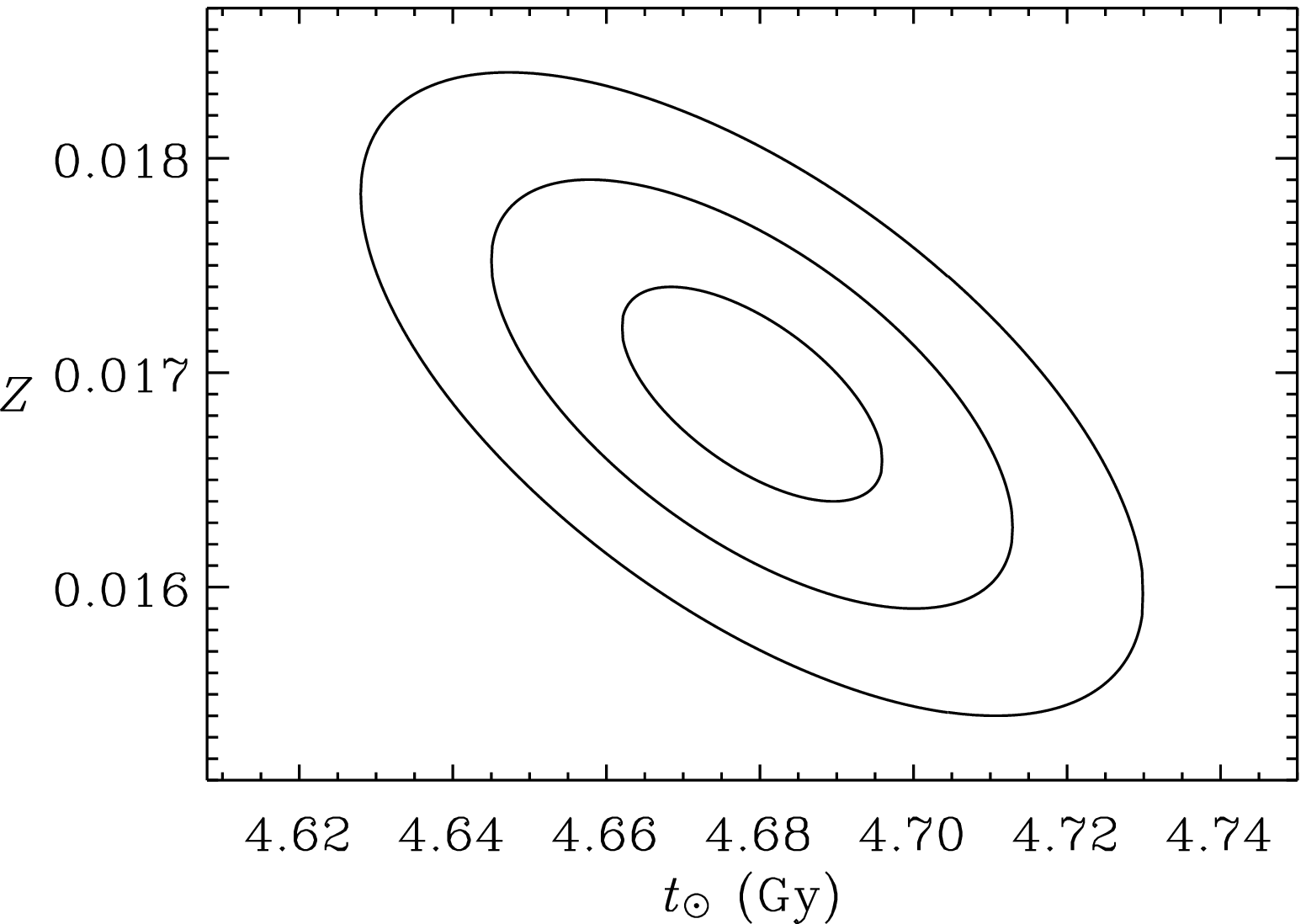,height=.20\textheight}
\end{minipage}
\hspace{2mm}
\begin{minipage}[h]{80mm}
  \vspace{-24mm} 
  \baselineskip=0.85\normalbaselineskip
  {\AIPtablecaptiontextfont{\bf FIGURE 4.\,\ }
   Error ellipses for the calibration using all three parameters $\xi_\alpha$: 
   solutions $(t_\odot,Z)$ satisfying the frequency data within 1, 2 and 3 
   standard errors in those data reside in the inner, intermediate 
   and outer ellipses, respectively.          
  }
\end{minipage}
}
\caption{\vspace{-8mm}}
\end{figure}
\renewcommand\AIPfigurecaptionheadformat[1] {\figurename\ #1.\hspace{1em}}

%
We thank J\o rgen Christensen-Dalsgaard for providing us with his 
stellar-evolutionary programme and for the instructions how to use it.
GH acknowledges support by the Particle Physics and Astronomy Research
Council of the UK.
%
%
\vspace{-4mm}
\bibliographystyle{mn2e}

\end{document}